\documentclass[pra,aps,aps,twocolumn,superscriptaddress]{revtex4}
\usepackage{amsfonts}
\usepackage{amsmath}
\usepackage{amssymb}
\usepackage{graphicx}
\usepackage{txfonts}

\begin{document}

\title{Multi-outcome homodyne detection in a coherent-state light
interferometer}
\author{J. Z. Wang}
\author{Z. Q. Yang}
\author{A. X. Chen}
\email{aixichen@zstu.edu.cn}
\affiliation{Department of Physics, Zhejiang Sci-Tech University, Hangzhou 310018, China}
\author{W. Yang}
\email{wenyang@csrc.ac.cn}
\affiliation{Beijing Computational Science Research Center, Beijing 100084, China}
\author{G. R. Jin}
\email{grjin@zstu.edu.cn}
\affiliation{Department of Physics, Zhejiang Sci-Tech University, Hangzhou 310018, China}
\date{\today }

\begin{abstract}
The Cram\'{e}r-Rao bound plays a central role in both classical and quantum
parameter estimation, but finding the observable and the resulting inversion
estimator that saturates this bound remains an open issue for general
multi-outcome measurements. Here we consider multi-outcome homodyne
detection in coherent-light Mach-Zehnder interferometer and construct a
family of inversion estimators that almost saturate the Cram\'{e}r-Rao bound
over the whole range of phase interval. This provides a clue on constructing
optimal inversion estimators for phase estimation and other parameter
estimation in any multi-outcome measurement.
\end{abstract}

\maketitle

\section{Introduction}

Interference of light fields is important in astronomy~\cite{Born,Caves},
spectroscopy~\cite{Wineland,Leibfried}, and various fields of quantum
technology~\cite{Dowling}. For instance, in optical lithography~\cite{Boto},
the light-intensity measurement gives rise to an oscillatory interferometric
signal $\propto\sin^{2}(\phi/2)$ or $\cos^{2}(\phi/2)$, which exhibits the
fringe resolution $\lambda/2$, determined by the wavelength $\lambda$. This
is often referred to the classical resolution limit of interferometer, or
the Rayleigh resolution criterion in optical imaging~\cite{Boto}. To beat
this classical limit, it has been proposed to use $N$-photon entangled
states $(|N, 0\rangle+|0, N\rangle)/\sqrt{2}$~\cite%
{Boto,Mitchell,Walther,Chen,Afek}, which results in the resolution $%
\lambda/(2N)$, i.e., $N$-times improvement beyond the classical resolution
limit. However, the $N$-photon entangled states are difficult to prepare and
are subject to photon losses~\cite{Braun,Dorner,Zhang}. In the absence of
quantum entanglement, the super-resolution can be also attainable from
different types of measurement schemes, such as coincidence photon counting~%
\cite{Resch}, parity detection~\cite{Gao,Cohen}, and homodyne detection with
post-processing~\cite{Andersen}.

To realize a high-precision measurement of an unknown phase $\phi $, an
optimal measurement scheme with a proper choice of data processing is
important to improve both the resolution and the phase sensitivity~\cite%
{Bevington,Dowling}. Given a phase-encoded state and a properly chosen
observable $\hat{\Pi}$, the ultimate phase estimation precision is
determined by the Cram\'{e}r-Rao lower bound (CRB)~\cite%
{Helstrom,Braunstein,Braunstein2,Paris1}, i.e., $\delta \phi \geq \delta
\phi _{\mathrm{CRB}}=1/\sqrt{F(\phi )}$, where $F(\phi )$ is the classical
Fisher information (CFI), dependent on the measurement probabilities. To
saturate the CRB, it requires complicated data-processing techniques such as
maximal likelihood estimation or Bayesian estimation~\cite{YMK,Pezze}.
However, they lack physical transparency and require a lot of computational
resources. By contrast, the simplest data processing is to equate the
theoretical expectation value $\langle \hat{\Pi}\rangle _{\phi }=f(\phi )$
with the experimentally measured value $\Pi _{\exp }$, which gives a simple
inversion estimator $\phi _{\mathrm{inv}}\equiv f^{-1}(\Pi _{\exp })$ to the
unknown phase $\phi $, with a precision determined by the error-propagation
formula:
\begin{equation}
\delta \phi =\frac{\Delta \hat{\Pi}}{\left\vert \partial \langle \hat{\Pi}%
\rangle _{\phi }/\partial \phi \right\vert }.  \label{EPF}
\end{equation}%
Here $\Delta \hat{\Pi}\equiv (\langle \hat{\Pi}^{2}\rangle _{\phi }-\langle
\hat{\Pi}\rangle _{\phi }^{2})^{1/2}$ is the root-mean-square fluctuation of
the observable $\hat{\Pi}$. Due to its physical transparency and
computational simplicity, the inversion estimator has been widely used in
various phase estimation schemes. Moreover, for binary-outcome measurements
(such as parity detection~\cite%
{Bollinger,Gerry2,Gerry2a,Gerry2b,Anisimov,Chiruvelli,Seshadreesan},
single-photon detection~\cite{Cohen}, and on-off measurement), the inversion
estimator asymptotically approaches the maximum likelihood estimator~\cite%
{Feng,Paris,Jin} and hence saturates the CRB within the whole phase
interval, e.g., $\phi \in \lbrack -\pi ,\pi ]$. However, for general
multi-outcome measurements, the performance of the inversion estimator
depends strongly on the chosen observable $\hat{\Pi}$ and usually cannot
saturate the CRB~\cite{Andersen}. Therefore, an important problem in
high-precision phase estimation is to find an optimal inversion estimator
that saturates the CRB. At present, this problem remains open.

In this paper, we investigate the performance of the inversion estimator in
multi-outcome measurements and report some interesting findings. We begin
with a binary observable, i.e., a multi-outcome measurement with only two
different eigenvalues. Such a binary observable plays a key role in
achieving deterministic super-resolution in a Mach-Zehnder interferometer
fed by a coherent-state light in a recent experiment~\cite{Andersen}. We
find that using a binary observable to construct the inversion estimator is
equivalent to binarizing the original multi-outcome measurement into an
effective binary-outcome one. As a result, the precision of the inversion
estimator \textit{saturates} the CRB determined by the CFI of this
binary-outcome measurement and is \textit{independent} of the choices of the
eigenvalues. We show that the observable adopted in Ref.~\cite{Andersen}
belongs to this class. Next, we consider the homodyne detection of Ref.~\cite%
{Andersen} as a paradigmatic example to study the dependence of the
precision of the inversion estimator on the observable. Surprisingly, we
find that when the neighboring eigenvalues of the observable have
alternating signs, the resulting inversion estimator is nearly optimal,
i.e., its precision \textit{almost} saturates the CRB. This may provide a
clue on constructing optimal inversion estimators for phase estimation and
other parameter estimation (e.g., optical angular displacements~\cite%
{Zhao,Cen,Wang,Xu,Qiang}).

\section{Quantum phase estimation with a multi-outcome measurement}

For a $(N+1)$-outcome measurement, the most general observable is
\begin{equation}
\hat{\Pi}=\sum_{k=0}^{N}\mu _{k}\hat{\Pi}_{k},  \label{observable}
\end{equation}%
where $\mu _{k}$ ($\hat{\Pi}_{k}$) is the eigenvalue (projector) associated
with the $k$th outcome. The output signal is the average of the observable $%
\hat{\Pi}$ with respect to the phase-encoded state $\hat{\rho}(\phi )$:
\begin{equation}
\langle \hat{\Pi}\rangle _{\phi }=\sum_{k=0}^{N}\mu _{k}P(k|\phi )\approx
\sum_{k=0}^{N}\mu _{k}\frac{{\mathcal{N}}_{k}}{{\mathcal{N}}},
\label{signal}
\end{equation}%
where $P(k|\phi )=\mathrm{Tr}[\hat{\rho}(\phi )\hat{\Pi}_{k}]$ is the
conditional probability of the $k$th outcome, which can be measured by the
occurrence frequency ${\mathcal{N}}_{k}/{\mathcal{N}}$. Specially, one can
perform ${\mathcal{N}}$ independent measurements at each phase shift $\phi
\in \lbrack -\pi ,\pi ]$ and record the occurrence numbers $\{{\mathcal{N}}%
_{k}\}$. For large enough ${\mathcal{N}}$, $P(k|\phi )\approx {\mathcal{N}}%
_{k}/{\mathcal{N}}$. Numerical simulation of a special kind of multi-outcome
measurement will be shown at the end of this work.

We begin with a binary observable -- the multi-outcome measurement with only
two different eigenvalues. Without losing generality, we take $\mu
_{N}=\mu_{N-1}=\cdots =\mu _{1}$, then we can use $\hat{\Pi}_{k}\hat{\Pi}%
_{k^{\prime}}=\delta _{kk^{\prime }}\hat{\Pi}_{k}$ to obtain
\begin{eqnarray*}
\langle \hat{\Pi}\rangle _{\phi } &=&\mu _{0}P(0|\phi )+\mu _{1}P(\emptyset
|\phi ), \\
\langle \hat{\Pi}^{2}\rangle _{\phi } &=&\mu _{0}^{2}P(0|\phi )+\mu
_{1}^{2}P(\emptyset |\phi ),
\end{eqnarray*}
where $P(\emptyset |\phi )\equiv \sum_{k=1}^{N}P(k|\phi )$ and $%
P(0|\phi)=1-P(\emptyset |\phi )$. Therefore, the phase sensitivity of the
inversion estimator based on the binary observable $\hat{\Pi}$ is
independent of the eigenvalues $\mu _{0}$ and $\mu _{1}$:
\begin{equation}
(\delta \phi )^{2}=\frac{P(\emptyset |\phi )P(0|\phi )}{\left[ P^{\prime
}(\emptyset |\phi )\right] ^{2}},  \label{sensitivity}
\end{equation}
where $P^{\prime }\equiv \partial P/\partial \phi $. Interestingly, if we
binarize the $N+1$ outcomes into two outcomes $0$ and \textquotedblleft $%
\emptyset $\textquotedblright, i.e., we regard the outcomes $1, 2, \cdots, N$
as a single outcome \textquotedblleft $\emptyset $\textquotedblright , then $%
P(\emptyset |\phi )$ is just the conditional probability for
\textquotedblleft $\emptyset $\textquotedblright\ and the CFI of this
effective binary measurement coincides with $1/(\delta \phi )^{2}$. In other
words, the phase sensitivity $\delta \phi $ of the inversion estimator based
on the binary observable $\hat{\Pi}$ always saturates the CRB of the
effective binary-outcome measurement~\cite{Feng}. Since this effective
binary-outcome measurement is obtained from the original multi-outcome
measurement by coarse-graining, its CFI is smaller than the CFI of the
original one:
\begin{equation}
F(\phi )\equiv \sum_{k=0}^{N}\frac{\left[ P^{\prime }(k|\phi )\right] ^{2}}{%
P(k|\phi )}.  \label{CFI}
\end{equation}
As a result, $\delta \phi $ cannot saturate the CRB of the original
multi-outcome measurement $\delta \phi _{\mathrm{CRB}}\equiv 1/\sqrt{F(\phi )%
}$ \cite{Helstrom,Braunstein,Braunstein2,Paris1}. Therefore, it is important
to find an optimal choice of eigenvalues $\{\mu _{k}\}$ such that $\delta
\phi =\delta\phi _{\mathrm{CRB}}$.

The above results are applicable to an arbitrary measurement scheme and
arbitrary input state. In the following, we consider the homodyne detection
at one port of Mach-Zehnder interferometer with a coherent-state input~\cite%
{Andersen} as a paradigmatic example to illustrate these results.
Interestingly, we find a nearly optimal observable that almost saturates the
CRB for all $\phi \in \lbrack -\pi ,\pi ]$.

\begin{figure}[htbp]
\begin{centering}
\includegraphics[width=1\columnwidth]{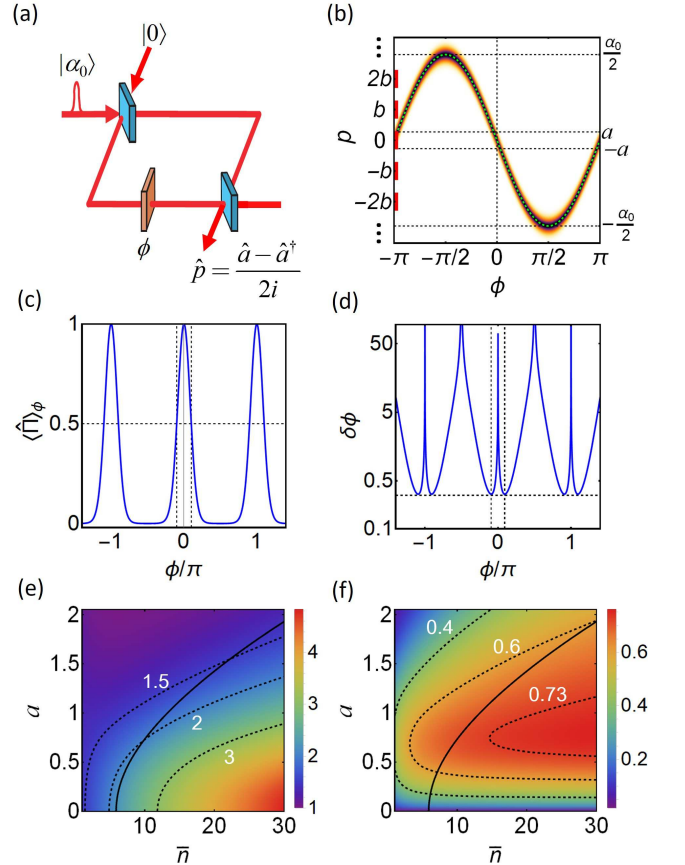}
\caption{(a) Homodyne detection at one port of the coherent-state
interferometer, equivalent to measuring quadrature operator $\hat{p}$ with
respect to the output state. (b) Conditional probability $P(p|\protect\phi)$
against the phase shift $\protect\phi$ and the measured quadrature $p$,
given by Eq.~(\protect\ref{CP}), and the post-processing method by
separating the measured data into several bins~\protect\cite{Andersen},
where the bin's center is $kb $ (for $k=0$, $\pm1$, ...$\pm k_{f}$) and the
width is $2a$, equivalent to a multi-outcome measurement. (c) and (d):
Output signal $\langle\hat{\Pi}\rangle_{\protect\phi}$ and phase sensitivity
$\protect\delta\protect\phi$ for $k_{f}=0$, equivalent to a binary-outcome
measurement. The vertical lines determine the resolution and the best
sensitivity, measured respectively by the full width at half maximum ($\mathrm{FWHM}$) and the best sensitivity $\delta\phi_{\min}$. (e) and (f): Density plots of the ratios $\frac {2\protect\pi/3}{\mathrm{FWHM}}$ and $\frac{1/\protect\sqrt{\bar{n}}}{\protect\delta\protect\phi_{\min}}$ as functions of the average photon number $\bar{n}$ $(=\protect\alpha_{0}^{2})$ and the bin size $a$. Dashed lines: contours of the two ratios. Solid lines and below: a region that the visibility of the signal $\geq 90\%$. }
\label{fig1}
\end{centering}
\end{figure}

As depicted by Fig.~\ref{fig1}(a), a coherent state $|\alpha _{0}\rangle $
and a vacuum state $|0\rangle $ are injected from the input ports. The
output state is given by $|\psi _{\mathrm{out}}(\phi )\rangle =\hat{U}(\phi
)|\psi _{\mathrm{in}}\rangle $, where $\hat{U}(\phi )$ is an unitary
operator
\begin{equation}
\hat{U}(\phi )=\exp \left( -i\frac{\pi }{2}\hat{J}_{y}\right) \exp \left(
-i\phi \hat{a}^{\dag }\hat{a}\right) \exp \left( -i\frac{\pi }{2}\hat{J}%
_{y}\right) ,  \label{U}
\end{equation}%
which represents a sequence actions of the 50:50 beamsplitter at the output
port~\cite{Gerrybook}, the phase accumulation at one of the two paths, and
the 50:50 beamsplitter at the input port. For brevity, we have introduced
Schwinger's representation of the angular momentum $\hat{J}=\frac{1}{2}(\hat{%
a}^{\dag },\hat{b}^{\dag })\hat{\sigma}\binom{\hat{a}}{\hat{b}}$, where $%
\hat{a}^{\dag }$ ($\hat{a}$) and $\hat{b}^{\dag }$ ($\hat{b}$) denote the
creation (annihilation) operators of the two field modes and $\hat{\sigma}=(%
\hat{\sigma}_{x},\hat{\sigma}_{y},\hat{\sigma}_{z})$ the Pauli matrix.

In general, a homodyne detection at one of two output ports gives the
measured quadrature $p\in (-\infty ,\infty )$, with the conditional
probability
\begin{equation}
P(p|\phi )=\int_{-\infty }^{\infty }dx\int_{-\infty }^{\infty
}dX\int_{-\infty }^{\infty }dPW_{\mathrm{out}}(\alpha ,\beta ;\phi ),
\end{equation}%
where $\alpha =x+ip$ ($x$, $y\in \mathbb{R}$) and $\beta =X+iP$ ($X$, $P\in
\mathbb{R}$), and $W_{\mathrm{out}}$ is the Wigner function of the output
state~\cite{Seshadreesan2,Tan},
\begin{equation}
W_{\mathrm{out}}(\alpha ,\beta ;\phi )=W_{\mathrm{in}}(\tilde{\alpha}_{\phi
},\tilde{\beta}_{\phi }),  \label{WignerOUT}
\end{equation}%
with
\begin{equation}
\left\{
\begin{array}{l}
\tilde{\alpha}_{\phi }=\alpha \frac{e^{i\phi }-1}{2}+\beta \frac{e^{i\phi }+1%
}{2}, \\
\tilde{\beta}_{\phi }=-\alpha \frac{e^{i\phi }+1}{2}-\beta \frac{e^{i\phi }-1%
}{2}.%
\end{array}%
\right.   \label{ab}
\end{equation}%
Note that Eqs.~(\ref{WignerOUT}) and (\ref{ab}) are valid for the two-path
interferometer described by $\hat{U}(\phi )$, independent from the input
state and the measurement scheme.

For the coherent-state input $|\psi _{\mathrm{in}}\rangle =|\alpha
_{0}\rangle \otimes |0\rangle $, the Wigner function is given by~\cite%
{Gerrybook}
\begin{equation}
W_{\mathrm{in}}(\alpha ,\beta )=\left( \frac{2}{\pi }\right)
^{2}e^{-2|\alpha -\alpha _{0}|^{2}}e^{-2|\beta |^{2}}.
\end{equation}%
Replacing $(\alpha ,\beta )$ by $(\tilde{\alpha}_{\phi },\tilde{\beta}_{\phi
})$, one can obtain the Wigner function of the output state, as shown by
Eq.~(\ref{WignerOUT}). For $\alpha _{0}=\sqrt{\bar{n}}\in \mathbb{R}$, we
obtain the probability to detecting an outcome $p$,
\begin{equation}
P(p|\phi )=\sqrt{\frac{2}{\pi }}\exp \left[ -2\left( p+\frac{\alpha _{0}}{2}%
\sin \phi \right) ^{2}\right] ,  \label{CP}
\end{equation}%
in agreement with our previous result~\cite{Feng}. In Fig.~\ref{fig1}(b), we
show density plot of $P(p|\phi )$ against the phase shift $\phi $ and the
measured quadrature $p$. The green dashed line is given by the equation $%
p=-\alpha _{0}\sin (\phi )/2$, indicating the peak of $P(p|\phi )$. The
commonly used observable in a traditional homodyne measurement is given by $%
\hat{\Pi}=\int_{-\infty }^{\infty }\!p|p\rangle \langle p|dp$, where $\hat{p}%
|p\rangle =p|p\rangle $ and $\hat{p}\equiv (\hat{a}-\hat{a}^{\dagger })/(2i)$%
. The output signal is the average of this observable $\langle \hat{\Pi}%
\rangle _{\phi }=\int_{-\infty }^{\infty }\!dppP(p|\phi )\propto \sqrt{\bar{n%
}}\sin \phi $, which exhibits the full width at half maximum ($\mathrm{FWHM}$%
)$\ 2\pi /3$, and hence the Rayleigh limit in fringe resolution. The
resolution can be improved by choosing a suitable observable~\cite{Andersen}.

\section{Binary-outcome homodyne detection}

We first consider the binary-outcome homodyne detection, where the measured
data has been divided into two bins~\cite{Andersen}: $p\in \lbrack -a,a]$ as
an outcome, denoted by \textquotedblleft $+$", and $p\notin \lbrack -a,a]$
as an another outcome \textquotedblleft $-$", with the bin size $2a$. Using
Eq.~(\ref{CP}), it is easy to obtain the conditional probabilities of the
outcomes \textquotedblleft $\pm $", namely
\begin{equation}
P(+|\phi )=\int_{-a}^{+a}dpP(p|\phi )=\frac{1}{2}\mathrm{Erf}\left[
g_{-}(\phi ),g_{+}(\phi )\right] ,  \label{Pplus_cs}
\end{equation}%
and hence $P(-|\phi )=1-P(+|\phi )$. Here, $\mathrm{Erf}\left[ x,y\right] =%
\mathrm{erf}(y)-\mathrm{erf}(x)$ denotes a generalized error function, and
\begin{equation}
g_{\pm }(\phi )=\sqrt{2}\left( \frac{\alpha _{0}}{2}\sin \phi \pm a\right) .
\label{gphi}
\end{equation}%
Obviously, this is a binary-outcome measurement with the observable $\hat{\Pi%
}=\mu _{+}\hat{\Pi}_{+}+\mu _{-}\hat{\Pi}_{-}$, where $\hat{\Pi}%
_{+}=\int_{-a}^{+a}|p\rangle \langle p|dp$ and $\hat{\Pi}_{-}=\hat{1}-\hat{%
\Pi}_{+}$. The signal is the average of $\hat{\Pi}$
\begin{equation}
\left\langle \hat{\Pi}\right\rangle _{\phi }=\mu _{+}P(+|\phi )+\mu
_{-}P(-|\phi ),  \label{sinal2}
\end{equation}%
where $\langle \hat{\Pi}_{\pm }\rangle _{\phi }=P(\pm |\phi )$. According to
Eq.~(\ref{sensitivity}), we obtain the phase sensitivity of the inversion
estimator:
\begin{equation}
\delta \phi =\frac{\sqrt{P(+|\phi )P(-|\phi )}}{\left\vert P^{\prime
}(+|\phi )\right\vert },
\end{equation}%
which is independent of the eigenvalues $\mu _{\pm }$ of the binary
observable $\hat{\Pi}$. The CFI of this binary-outcome measurement is given
by~\cite{Feng,Paris,Jin}
\begin{equation}
F(\phi )=\sum_{k=\pm }\frac{\left[ P^{\prime }(k|\phi )\right] ^{2}}{%
P(k|\phi )}=(\delta \phi )^{-2},
\end{equation}%
where, in the last step, we have used the relation $P(+|\phi )+P(-|\phi )=1$%
. Therefore, the sensitivity $\delta \phi $ of the inversion estimator based
on the binary observable $\hat{\Pi}$ always saturates the CRB of the
binary-outcome measurement~\cite{Feng,Paris,Jin}.

In Figs.~\ref{fig1}(c) and \ref{fig1}(d), we take $\mu _{+}=1/\mathrm{erf}(%
\sqrt{2}a)$ and $\mu _{-}=0$ to show the signal and the sensitivity\ as
functions of $\phi $, where $\mathrm{erf}(\sqrt{2}a)$ is a normalization
factor~\cite{Andersen}. The vertical lines determine the resolution and the
best sensitivity $\delta \phi _{\min }=1/\sqrt{F(\phi _{\min })}$. From
Figs.~\ref{fig1}(e) and \ref{fig1}(f), one can find that a higher resolution
with the $\mathrm{FWHM}\sim \pi /\sqrt{\bar{n}}$, can be obtained as the bin
size $a\rightarrow 0$. However, the best sensitivity occurs as $a\geq 1/2$.
Therefore, as a trade-off, one can simply take $a=(\Delta \hat{p}%
)_{|\alpha\rangle }=1/2$~\cite{Andersen}, for which both the resolution and
the sensitivity scale inversely with $\sqrt{\bar{n}}$. Numerically, it has
been shown that the best sensitivity can reach $\delta \phi _{\min }\sim
1.37/\sqrt{\bar{n}}$~\cite{Andersen}.

We now investigate the visibility of the interferometric signal and its
relationship with $a$ and $\bar{n}$ that have not been addressed by Ref.~%
\cite{Andersen}. From Fig.\ref{fig1}(c), one can note that the visibility
can be determined by
\begin{equation}
V=\frac{\left\langle \hat{\Pi}\right\rangle _{\phi =0}-\left\langle \hat{\Pi}%
\right\rangle _{\phi =\pi /2}}{\left\langle \hat{\Pi}\right\rangle _{\phi
=0}+\left\langle \hat{\Pi}\right\rangle _{\phi =\pi /2}},  \label{visibility}
\end{equation}
where $\phi =\pm \pi /2$ denote the dark points of the signal. Using Eqs.~(%
\ref{Pplus_cs}) and (\ref{sinal2}), we can easily obtain an analytical
result of the visibility, which gives a relation between $a$ and $\bar{n}$
for a given $V$. The solid line of Fig.~\ref{fig1}(e) corresponds to $V=0.9$
and a region below it indicates $V>0.9$. Our numerical results show that the
visibility is larger than $90\%$ only when the average number of photons is
not too small (at least $\bar{n}>5.8$).

\section{Multi-outcome homodyne detection}

To proceed, let us consider the multi-outcome case by separating the
measured quadrature into several bins~\cite{Andersen}, i.e., treating $p\in
\lbrack b_{k}-a,b_{k}+a]$ as an outcome $k$, where $b_{k}$ is center of each
bin. When $p$ does not lie within any bin, then we identify $p$ as outcome
\textquotedblleft $-$\textquotedblright . The occurrence probability of the $%
k$-th outcome is
\begin{align}
P(k|\phi )& =\int_{b_{k}-a}^{b_{k}+a}dpP(p|\phi )  \notag \\
& =\frac{1}{2}\mathrm{Erf}\left[ g_{-}(\phi )+\sqrt{2}b_{k},g_{+}(\phi )+%
\sqrt{2}b_{k}\right] ,  \label{Pbk}
\end{align}%
where $g_{\pm }(\phi )$ is defined in Eq.~(\ref{gphi}). The occurrence
probability for the outcome \textquotedblleft $-$" is simply given by $%
P(-|\phi )=1-\sum_{k}P(k|\phi )$. According to Distante \textit{et al.}~\cite%
{Andersen}, one can take $b_{k}=kb$ with the integers $k=0$, $\pm 1$, $%
\cdots $, $\pm k_{f}$ and a factor $b$ to be determined, so the total number
of the outcomes is $2(k_{f}+1)$. For a given $b>2a$, the overlap of the
conditional probabilities between neighbour outcomes is vanishing, and $%
k_{f}b\sim \alpha _{0}/2$ [see Fig.~\ref{fig1}(b)]. For large enough $\bar{n}%
=\alpha _{0}^{2}$ and $b$ ($>2a$), we take $k_{f}=[\alpha _{0}/(2b)]$ such
that the outcomes for $|k|>k_{f}$ give almost vanishing contribution, where $%
[x]$ is the integer closest to $x$.

\begin{figure}[tbph]
\begin{centering}
\includegraphics[width=1\columnwidth]{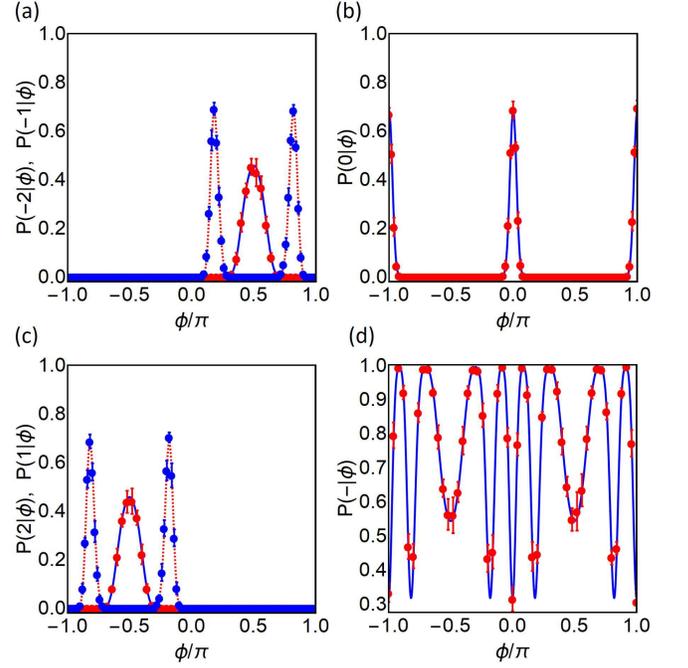}
\caption{Conditional probabilities $P(k|\phi)$ for $k=0$, $\pm 1$, ..., $\pm k_f$, and $P(-|\phi)=1-\sum_{k}P(k|\phi)$. The parameters: $\bar{n}=200$, $a=1/2$, $b=3.8$, and hence the total number of the outcomes $2(k_f+1)=6$, since $k_f=2$ (see text). Left panel: $P(\pm 2|\phi)$ for the blue solid lines, and $P(\pm 1|\phi)$ for the red dashed lines. Numerical simulations: averaged occurrence frequency ${\mathcal N}_k/{\mathcal N}$ (the solid circles) and its standard derivation (the bars) of each outcome after $M=10$ replicas of ${\mathcal N}=200$ independent measurements.}
\label{fig2}
\end{centering}
\end{figure}

In Fig.~\ref{fig2}, we take $\alpha _{0}=\sqrt{200}$, $a=1/2$, and $b=3.8$
to show the occurrence probabilities as function of $\phi$. Here, the total
number of the outcomes $2(k_{f}+1)=6$, since $k_{f}=2$. The visibility of $%
P(0|\phi)$ is similar to the binary-outcome problem and is larger than $90\%$
as long as $\bar{n}=\alpha _{0}^2>5.8$. Numerical simulations of the
occurrence probabilities are shown using random numbers ranged from $0$ to $%
1 $ (see below).

\begin{figure}[ptbh]
\begin{centering}
\includegraphics[width=1\columnwidth]{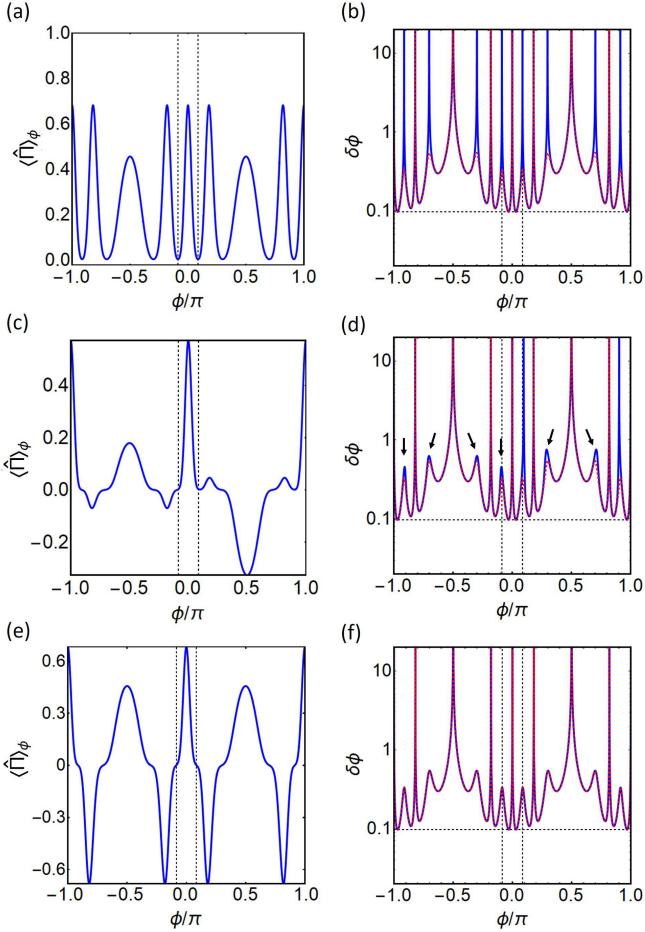}
\caption{Output signal $\langle\hat{\Pi}\rangle_{\phi}$ and phase sensitivity $\delta\phi$ for $\mu_{-}=0$ and different choices of $\{\mu_{k}\}$. The parameters: $\bar{n}=200$, $a=1/2$, $b=3.8$, and $k_f=2$. (a) and (b): $\mu_{k}=1$ for all $k$'s. (c) and (d): $\{\mu_{-2}, \mu_{-1}, \mu_{0}, \mu_{1}, \mu_{2}\}=\{-0.715, 0.068, 0.839, -0.102, 0.392\}$. (e) and (f): $\{\mu_{-2}, \mu_{-1}, \mu_{0}, \mu_{1}, \mu_{2}\}=\{1, -1, 1, -1, 1\}$. The red dashed lines: the CRB $1/\sqrt{F(%
\phi )}$. The vertical lines: locations of the first dark points $\phi _{\mathrm{dark}}\approx \pm b/\sqrt{\bar{n}}$. The horizontal lines in right panel: the best sensitivity $\delta\phi_{\min}\approx
1.37/\sqrt{\bar{n}}$.}
\label{fig3}
\end{centering}
\end{figure}

The observable corresponding to this multi-outcome homodyne detection can be
written as $\hat{\Pi}=\mu _{-}\hat{\Pi}_{-}+\sum_{k}\mu _{k}\hat{\Pi}_{k}$,
where $\hat{\Pi}_{k}\equiv \int_{b_{k}-a}^{b_{k}+a}|p\rangle \langle p|dp$
and $\hat{\Pi}_{-}\equiv \hat{1}-\sum_{k}\hat{\Pi}_{k}$, with the
eigenvalues $\{\mu _{k}\}$ and $\mu _{-}$. The signal is the average of $%
\hat{\Pi}$:
\begin{equation}
\left\langle \hat{\Pi}\right\rangle _{\phi }=\sum_{k=-k_{f}}^{k_{f}}\mu
_{k}P(k|\phi )+\mu _{-}P(-|\phi ),  \label{signal3}
\end{equation}
Previously, Distante et al.~\cite{Andersen} have derived the signal and the
sensitivity by taking $\mu _{k}=1/\mathrm{erf}(\sqrt{2}a)$\ and $\mu _{-}=0$%
. For arbitrary $\mu _{k}=\mu _{+}$ and $\mu _{-}=0$, it is easy to reduce
the signal as $\langle \hat{\Pi}\rangle _{\phi }=\mu _{+}-(\mu _{+}-\mu
_{-})P(-|\phi )$. Furthermore, the sensitivity is found independent on the
values of $\mu _{\pm }$; see Eq.~(\ref{sensitivity}). In Figs.~\ref{fig3}(a)
and \ref{fig3}(b), we simply take $\mu _{k}=\mu _{+}=1$ and $\mu _{-}=0$ to
show the signal $\langle \hat{\Pi}\rangle _{\phi }=1-P(-|\phi )$ and the
sensitivity against $\phi $. Similar to Ref.~\cite{Andersen}, the signal
exhibits a multi-fold oscillatory pattern, with the peaks located at
\begin{equation}
\phi _{k}=\arcsin \left( \frac{2b_{k}}{\alpha _{0}}\right) ,  \label{peaks}
\end{equation}
and also $\pi -\phi _{k}$. If the bin's center $b_{k}=kb$ and $|k|<k_{f}$,
it is easy to obtain $\phi _{k}\approx 2kb/\alpha _{0}$ and hence the first
dark point of the signal $\phi _{\mathrm{dark}}\approx \phi _{\pm
1}/2\approx \pm b/\alpha _{0}$; see the vertical lines of Fig.~\ref{fig3}(a).

When all $\{\mu _{k}\}$ are the same, we obtain a binary observable $\hat{\Pi%
}$ and the sensitivity is similar to Eq.~(\ref{sensitivity}),
\begin{equation}
\delta \phi =\frac{\sqrt{P(+|\phi )P(-|\phi )}}{|P^{\prime }(-|\phi )|}\geq
\frac{1}{\sqrt{F(\phi )}},  \label{sensitivity3}
\end{equation}
where $P(+|\phi )\equiv \sum_{k}P(k|\phi )=1-P(-|\phi )$, and $F(\phi )$ is
the CFI of the multi-outcome homodyne measurement,
\begin{equation}
F(\phi )=\sum_{k=-k_{f}}^{k_{f}}\frac{\left[ P^{\prime }(k|\phi )\right]
^{2} }{P(k|\phi )}+\frac{\left[ P^{\prime }(-|\phi )\right] ^{2}}{P(-|\phi )}%
.  \label{CFI3}
\end{equation}
In Fig.~\ref{fig3}(b), we show the sensitivity $\delta \phi $ and its
ultimate lower bound $\delta \phi _{\mathrm{CRB}}=1/\sqrt{F(\phi )}$ against
$\phi $. One can see that the sensitivity $\delta \phi $ diverges at certain
values of $\phi $ (e.g., $\phi _{\mathrm{dark}}\approx \pm b/\alpha _{0}$),
but $\delta \phi _{\mathrm{CRB}}$ does not (see the red dashed line). The
singularity of $\delta\phi$ means that complete no phase information can be
inferred. Usually, it takes place when the slope of signal is vanishing,
i.e., $P^{\prime }(-|\phi )=0$. On the other hand, $\delta \phi _{\mathrm{CRB%
}}$ diverges when $F(\phi )=0$, i.e., $P^{\prime}(-|\phi )=P^{\prime
}(k|\phi )=0$ at certain values of $\phi$.

As depicted by Figs.~\ref{fig3}(a) and \ref{fig3}(b), the sensitivity $%
\delta \phi $ diverges at the extreme values of the signal; While for $%
\delta \phi _{\mathrm{CRB}}$, however, the divergences only occur at the
peaks of the signal. The reason why the sensitivity shows a series of \emph{%
extra} divergences at the minima of the output signal could be understood by
the fact that the signal is a sum of highly sharp phase distribution as Fig.~%
\ref{fig2}, weighted by positive eigenvalues $\mu _{k}=+1$. It is therefore
important to investigate the dependence of the signal and the sensitivity on
different choices of the eigenvalues, which has \textit{not} been addressed
in Ref.~\cite{Andersen}.

For a general multi-outcome measurement, different choices of $\{\mu _{k}\}$
correspond to different observables $\hat{\Pi}=\sum_{k}\mu _{k}\hat{\Pi}_{k}$%
. Therefore, both the signal and the sensitivity depend on the eigenvalues $%
\{\mu _{k}\}$. To see it clearly, we take random numbers of $\{\mu _{k}\}$,
ranged from $-1$ to $+1$. As shown by the blue solid lines of Figs.~\ref%
{fig3}(c) and \ref{fig3}(d), one can see that both the signal and the
sensitivity are different with that of Figs.~\ref{fig3}(a) and ~\ref{fig3}%
(b). However, near $\phi =0$, the sensitivity in Fig.~\ref{fig3}(d)
coincides with that of Fig.~\ref{fig3}(b), indicating that Eq.~(\ref%
{sensitivity3}) still works to predict the best sensitivity. Remarkably, one
can also see that the sensitivity does not diverge at the locations of the
arrows. The singularity of $\delta \phi $ can be further suppressed using
alternating signs of $\{\mu _{k}\}$ for neighbour outcomes (i.e., $\mu
_{k\pm 1}=-\mu _{k}$). As shown in Figs.~\ref{fig3}(e) and \ref{fig3}(f),
one can see $\delta \phi \approx \delta \phi _{\mathrm{CRB}}=1/\sqrt{F(\phi )%
}$ within the whole phase interval. In other words, the inversion estimator
associated with the so-chosen observable almost saturates the CRB.

\begin{figure}[hptb]
\begin{centering}
\includegraphics[width=1\columnwidth]{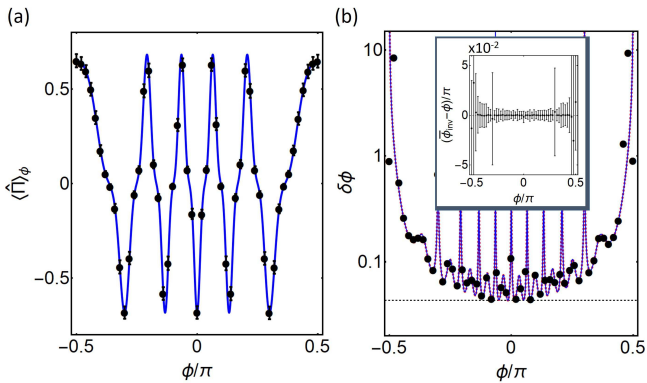}
\caption{Output signal $\langle\hat{\Pi}\rangle_{\phi}$ and phase sensitivity $\delta\phi$ for alternating signs of $\{\mu_{k}\}$ and the parameters: $\bar{n}=1000$, $a=1/2$, $b=3.2$, and $k_f=5$ (i.e., total number of the outcomes is $12$). Solid circles: the averaged signal and the phase uncertainty of the inversion estimator $\phi_{\mathrm{inv}}$ obtained with ${\mathcal N}=200$ and $M=400$. The horizontal line in (b): the best sensitivity $\delta\phi_{\min}\approx 1.37/\sqrt{\bar{n}}$. Inset: Difference between the average value of the inversion estimator $\bar{\phi}_{\mathrm{inv}}=\sum_{i=1}^{M}\phi_{\mathrm{inv}}^{(i)}/M$ and the true value of phase shift $\phi$. The bars are the standard deviations of the estimators $\{\phi_{\mathrm{inv}}^{(1)},\phi_{\mathrm{inv}}^{(2)}, ..., \phi_{\mathrm{inv}}^{(M)}\}$.}
\label{fig4}
\end{centering}
\end{figure}

We now adopt Monte Carlo method to simulate the above multi-outcome
measurement~\cite{Pezze,Jin}. Specially, we first generate $\mathcal{N}$
random numbers $\{\xi _{1},\xi _{2}, ...,\xi _{\mathcal{N}}\}$, according to
the occurrence probabilities $\{ P(k|\phi )\}$ at a given $\phi$, where $\xi
_{i}$ (for $i=1$, $2$, $\cdots$, $\mathcal{N}$) can be regarded as the
outcome $k=-k_{f}$, provided $0\leq \xi _{i}\leq P(-k_{f}|\phi)$. It can be
regarded as the outcome $k=-k_{f}+1$, provided $P(-k_{f}|\phi )<\xi _{i}\leq
P(-k_{f}|\phi )+P(-k_{f}+1|\phi )$, and so on. If $\xi _{i}$ obeys $%
P(-k_{f}|\phi )+P(-k_{f}+1|\phi )+\cdots +P(k_{f}|\phi )<\xi _{i}\leq 1$,
then we treat it as the outcome \textquotedblleft $-$". In this way, we
obtain the occurrence numbers\ of all the outcomes $\{\mathcal{N}_{k}\}$.
Next, we repeat the above process for any value of $\phi \in (-\pi ,\pi )$
and obtain the occurrence frequencies $\{\mathcal{N}_{k}/\mathcal{N}\}$. As
depicted by Fig.~\ref{fig2}, we show the averaged $\mathcal{N}_{k}/\mathcal{N%
}$ (the solid circles) and its standard deviation (the bars) after $M=10$
replicas of the above simulations. With large enough $\mathcal{N}$ ($=200$),
one can see that the averaged occurrence frequency of each outcome almost
follows its theoretical result, as is expected. Using Eqs.~(\ref{signal})
and (\ref{signal3}), we further obtain the average signal and its standard
deviation. In Fig.~\ref{fig4}(a), we take $\bar{n}=\alpha _{0}^{2}=1000$ and
$b=3.2$, which gives $k_{f}=5$ and hence the number of outcomes $12$.
Similar to Figs.~\ref{fig3}(e) and (f), we choose alternating signs of the
eigenvalues $\{\mu _{k}\}$ and vanishing $\mu _{-}$.

In real experiments, the dependence of $P(k|\phi )$ on $\phi $ is obtained
from replicas of $\mathcal{N}$ independent measurements at each given phase
shift. This comprises a calibration of the interferometer. With all known
occurrence probabilities and the signal, one can infer unknown value of $%
\phi $ via the phase estimation. As the simplest protocol, we adopt the
inversion phase estimator $\phi _{\mathrm{inv}}=g^{-1}(\sum_{k}\mu _{k}{%
\mathcal{N}_{k}}/{\mathcal{N}})$, where $g^{-1}$ is the inverse function of
the average signal $g(\phi )=\langle \hat{\Pi}\rangle _{\phi }$ and ${%
\mathcal{N}_{k}}/{\mathcal{N}}$ is the occurrence frequency of the $k$th
outcome in a single $\mathcal{N}$ independent measurements. After $M$
replicas, one can obtain the estimators $\{\phi _{\mathrm{inv}}^{(1)},\phi _{%
\mathrm{inv}}^{(2)},...,\phi _{\mathrm{inv}}^{(M)}\}$. The mean value of the
estimators $\bar{\phi}_{\mathrm{inv}}=\langle \phi _{\mathrm{inv}%
}^{(i)}\rangle _{s}$ and its standard deviation are shown in the inset of
Fig.~\ref{fig4}(b), where the statistical average is defined as $\langle
(\cdots )\rangle _{s}\equiv \sum_{i=1}^{M}(\cdots )/M$. The standard
deviation (the bars) is larger than $(\bar{\phi}_{\mathrm{inv}}-\phi )$
indicates that the inversion estimator is unbiased; see the inset of Fig.~%
\ref{fig4}(b). For an effective single-shot measurement, the phase
uncertainty is defined by $\sigma =\sqrt{{\mathcal{N}}}\sqrt{\langle (\phi _{%
\mathrm{inv}}^{(i)}-\phi )^{2}\rangle _{s}}$, which almost follows the lower
bound of phase sensitivity $\delta \phi _{\mathrm{CRB}}=1/\sqrt{F(\phi )}$;
see the solid circles of Fig.~\ref{fig4}(b).

Finally, it should be mentioned that we have discussed the achievable
sensitivity close to the shot-noise limit with the coherent-state input.
However, it is possible to surpass this classical limit once the
interferometer is fed by nonclassical states of light. Recently, Sch\"{a}%
fermeier \textit{et al}~\cite{Schafermeier} have demonstrated that both the
resolution and the sensitivity can surpass their classical limits using the
binary-outcome homodyne detection, where a coherent state and a squeezed
vacuum are used as the input. For any binary-outcome measurement, we have
shown that the phase estimator by inverting the average signal is good
enough to saturate the CRB~\cite{Feng,Paris,Jin}. For a multi-outcome
detection, the inversion estimator is less optimal due to the divergence of
the phase sensitivity~\cite{YMK,Pezze}. We show here that the singularity
can be suppressed when the signal is a sum of positive $P(k|\phi)$, weighted
by alternating signs of eigenvalues.

\section{Conclusion}

In summary, we have considered quantum phase estimation with multi-outcome
homodyne detection in the coherent-state light Mach-Zehnder interferometer.
Compared with the ultimate phase sensitivity determined by the classical
Fisher information, we show that (i) the phase sensitivity shows a series of
extra divergences at the minima of the output signal; (ii) these extra
divergences can be removed by using observables whose eigenvalues associated
with neighboring outcomes have alternating signs. This result provides a
family of nearly optimal inversion estimators that almost saturate the Cram%
\'{e}r-Rao bound over the whole range of phases. We further perform
numerical simulations using such observables and demonstrate that phase
uncertainty of the inversion estimator almost follows the Cram\'{e}r-Rao
bound. Our method for removing extra divergences of the phase sensitivity
may also be applicable to other kinds of multi-outcome measurements.

\section*{Funding}

Science Foundation of Zhejiang Sci-Tech University (18062145-Y); National
Natural Science Foundation of China (NSFC) (91636108, 11775190, and
11774021); the NSFC program for ``Scientific Research Center" (U1530401).

\section*{Acknowledgments}

We thank Professor C. P. Sun for helpful discussions.


\end{document}